\documentclass[structabstract]{aa}
\usepackage{graphicx}
\usepackage{txfonts}
\begin{document}

   \title{Modelling chromospheric line profiles as diagnostics of \\ 
       velocity fields in $\omega$ Centauri red giant stars}


\titlerunning{Chromospheres and mass motions in $\omega$ Cen red giant stars}

   \author{M. Vieytes\inst{1}, P. Mauas\inst{1}\thanks{Visiting Astronomer at the INAF Osservatorio 
   Astronomico Bologna, under bilateral agreement of Ministero degli Affari Esteri (Italy) 
   and Ministerio de Asuntos Exteriores (Argentina).}, C. Cacciari\inst{2}\thanks{Visiting Astronomer 
   at the IAFE Buenos Aires, under bilateral agreement of Ministero degli Affari Esteri (Italy) 
   and Ministerio de Asuntos Exteriores (Argentina).}, L. Origlia\inst{2} 
          \and
          E. Pancino\inst{2}\fnmsep
          }
	  
\authorrunning{M. Vieytes et al.} 

   \institute{Instituto de Astronom\'\i a y F\'\i sica del Espacio, Universidad de Buenos Aires, 
              Argentina\\
           \email{mariela@iafe.uba.ar, pablo@iafe.uba.ar}             
         \and
             INAF - Osservatorio Astronomico, Via Ranzani 1, 40127 Bologna, Italy\\
             \email{carla.cacciari@oabo.inaf.it, livia.origlia@oabo.inaf.it, elena.pancino@oabo.inaf.it}
             }
	     
       \date{Received 22/07/2010; accepted 24/10/2010}	     

\abstract
{Mass loss of $\sim$ 0.1-0.3 M$_{\odot}$ from Population II red giant stars (RGB) is a requirement 
of stellar evolution theory in order to account for several 
observational evidences in globular clusters. }
{The aim of this study is to detect the presence of outward velocity fields, which are 
indicative of mass outflow, in six luminous red giant stars of the stellar cluster $\omega$  Cen.}
{We compare synthetic line profiles computed using relevant model chromospheres 
to observed profiles of the H$\alpha$ and Ca II K lines.  
The spectra were taken with UVES (R=45,000) and the stars were selected so that three 
of them belong to the metal-rich population and three to the metal-poor population, 
and sample as far down as 1 to 2.5 magnitudes fainter than the respective RGB tips.}
{We do indeed reveal the presence of low-velocity outward motions in four of 
our six targets,  without any apparent correlation with astrophysical parameters.}
{This provides direct evidence that outward velocity fields and mass motions exist in 
RGB stars as much as 2.5 mag fainter than the tip. On the assumption that the mass 
outflow may eventually lead to mass loss from the star, we estimate mass-loss rates 
of some 10$^{-9}$-10$^{-10} M_{\odot}$ yr$^{-1}$ that are compatible with the 
stellar evolution requirements.  These rates seem to be correlated with 
luminosity rather than metallicity. 
}



   \maketitle
%

\section{Introduction}\label{s:int}

Mass loss of $\sim$ 0.1-0.3 M$_{\odot}$ from Population II red giant branch (RGB) 
stars is a requirement of stellar evolution theory in order to account for 
observational evidence in globular clusters (GCs) such as 
i) the very existence of the horizontal branch (HB) and its morphology, 
ii) the pulsational properties of RR Lyrae stars, 
iii) the absence of asymptotic giant branch (AGB) stars brighter than the red giant 
branch (RGB) tip, and the chemistry and characteristics in the AGB, post-AGB 
and planetary nebula evolutionary phases, 
iv) the mass of white dwarf (WD) stars (Castellani \& Renzini 1968; Rood 1973; Fusi Pecci
et al. 1993; D'Cruz et al. 1996; Kalirai et al. 2007). 
In addition, mass loss  significantly affects 
other areas of astrophysics, for example the UV excess in elliptical galaxies 
and the interaction between the cool intracluster medium and hot halo gas. 

The few tens of solar masses lost by RGB stars should accumulate in the 
central regions of GCs, especially of the more massive ones, without the  
sweeping mechanisms between Galactic plane crossings. 
However, no conclusive evidence was found for a significant 
presence of any form (neutral or ionized gas, dust) of intracluster matter, 
even in some of the clusters 
that would be most likely to accumulate the material lost from cluster 
stars (Smith et al. 1990; Origlia et al. 1997; Freire et al. 2001).
The advent of the {\it Spitzer Space Telescope} has triggered a more accurate 
search for cold dust within GCs. Eight Galactic GCs were observed and again 
only upper limits were found, which is well below expectations for 
dust production from mass loss in cluster RGB and AGB stars 
(Barmby et al. 2009, and references therein).
The only  exception is NGC 7078 (M15) which  shows high signal-to-noise 
evidence for an IR excess and hence dusty intracluster medium
in the core (Evans et al. 2003; Boyer et al. 2006),
and possibly $\omega$ Cen if the four dust clouds that were detected in the field 
are confirmed to be intracluster material that is in the process 
of escaping from the cluster (Boyer et al. 2008).
This implies that either dust production from mass loss is less efficient, or that 
the mechanisms able to remove the intracluster material are more efficient than 
commonly believed. 

An alternative and complementary approach is to study the mass loss phenomenon where 
it occurs, namely on the RGB stars themselves, by detecting either outflow motions in 
the outer region of the stellar atmospheres or the presence of circumstellar envelopes 
at larger distances (typically tens/hundreds of stellar radii). 

The latter approach, which needs the use of IR data, was first made possible by mid-IR 
photometry taken from the ground (Frogel \& Elias 1988) 
and with the IRAS satellite (Gillet et al. 1988; Origlia et al. 1996),
and a decade later by new observations from the ISOCAM onboard ISO. 
A deep survey on six massive Galactic GCs with ISOCAM data revealed an  
infrared excess, which is indicative of dusty circumstellar envelopes, 
in $\sim$ 15\% of the RGB 
(or AGB) stars in the $\sim$ 0.7 mag brightest interval (M$_{\rm bol} \le -2.5$) 
(Origlia et al. 2002, and references therein).
A  more accurate survey in 17 Galactic GCs was made possible recently  by the 
use of data from the IRAC onboard {\it Spitzer} (Fabbri et al. 2008).
The results from this survey on about 100 stars in 47 Tuc (Origlia et al. 2007, 2010)
show unambiguously that mid-IR excess radiation caused by dust 
is present along the RGB down to about the level of the HB and is correlated with 
luminosity. 
Mass-loss rates were derived for these stars, allowing to define  
the first empirical mass-loss formula calibrated on Population II stars. 
The dependence on luminosity of this mass-loss rate is considerably shallower 
than the widely used Reimers law (Reimers 1975a,b)
and the formulation by Catelan (2000) 
which were both calibrated  on Population I red giants. 
We refer the reader to Origlia (2008) for a review.

Other {\it Spitzer}-based studies available on 47 Tuc (Boyer et al. 2010),
$\omega$ Cen (Boyer et al. 2008; Mc-Donald et al. 2009), 
M15 (Boyer et al. 2006),  
and NGC 362 (Boyer et al. 2009) 
find that mass loss is confined near the tip of the RGB, and occurs mostly in AGB or variable 
stars, with the exception of M15 where a population of dusty red giants has been detected. 
These conclusions, however, were revised by Origlia et al. (2010) 
in 47 Tuc where the use of a better dust indicator allowed them to detect  
dust at fainter magnitudes along the RGB.  

The other approach, aiming at detecting velocity fields and outward motions in the 
stellar chromospheres, has also been the subject of several studies since the late 70's. 
This is the approach we adopt in the present study, as described in detail in the 
following sections.

\section{Diagnostics of velocity fields}\label{s:diag}

The pioneering investigations of the H$\alpha$ line profile in globular cluster red giants 
(Cohen 1976,1978, 1979, 1980, 1981; Mallia \& Pagel 1978; Peterson 1981, 1982; Cacciari 
\& Freeman 1983; Gratton et. al 1984) 
detected asymmetric and variable H$\alpha$ emission wings along the brightest part of the 
RGB, which were generally interpreted as evidence of an extended atmosphere until 
Dupree et al. (1984) showed that they could instead be signatures of a static 
stellar chromosphere. 

Therefore, the attention shifted to  {\em asymmetry} and {\em coreshift} in 
chromospheric line profiles as possible indicators of mass motions. 
This approach was applied to 
the Na I D and Ca II K lines in addition to H$\alpha$, in both globular cluster RGB 
stars (Peterson 1981; Bates et al. 1990, 1993; Dupree et al. 1994; Lyons et al. 1996; 
Cacciari et al. 2004; M\'esz\'aros et al. 2008, 2009a) 
and in metal-poor field red giants (Smith et al. 1992; Dupree \& Smith 1995).
Blueshifted (outward) velocity core shifts were detected in several 
cases, all of them much smaller than the escape velocity from the stellar photosphere. 
Better results were obtained using the Mg II $\lambda$2800 $h$ and $k$ lines which form 
higher in the atmosphere. These data are only available  for few bright metal-poor 
field or globular cluster red giants (Dupree et al. 1990a,b, 1994; Smith \& Dupree 1998),
and indeed  allowed the detection of stellar winds with terminal speeds exceeding the
escape velocities, from  which mass-loss rates of about 
10$^{-11}$ to 10$^{-9}$ M$_{\odot}$ yr$^{-1}$ were estimated. 

The infrared He I $\lambda$10830 line, which forms still higher in the atmosphere 
than optical or near-UV chromospheric lines, has been proposed to  probe 
the outer regions of the atmosphere where the wind begins to accelerate (Dupree et al. 1992;
Smith et al. 2004). 
In a recent study of 41 metal-poor field giants, Dupree et al. (2009)
detected fast outflows of material, in about 40\% of cases fast enough 
to escape from the star.
From their model chromosphere and the analysis of line strength and outflow speed they 
derived mass loss rates in the range $\sim$ 10$^{-10}$ to 10$^{-8}$ M$_{\odot}$ yr$^{-1}$, 
which provide the amount of mass loss ($\sim$ 0.2 M$_{\odot}$) needed by stellar 
evolution models during the RGB and RHB phases. However, as they point
out, helium lines can be strongly affected by X-ray photoionization of
He, as was shown for the Sun by Mauas et al. (2005)
and Andretta et al. (2008).

In any case, this potentially interesting method can
be presently applied only to bright (i.e. field) stars, given the
difficulty of obtaining high-resolution high S/N spectra in the near IR. 
Therefore, the use of the optical spectroscopic diagnostics can still provide 
important information, and it may well be the only viable option to study the much 
fainter RGB stars in globular clusters, at least until powerful high-resolution 
IR spectrographs become widely available. 

The advantage of monitoring the mass-loss phenomenon in globular cluster RGB stars 
is the convenience of studying a sample with the same metallicity and distance  
within each cluster, and therefore the possibility to detect with relatively good 
accuracy the dependence of the mass-loss phenomenon on luminosity (within a given 
cluster) and on chemical abundance (by comparing results from different clusters). 
This was the aim of the most recent studies, which used detailed and custom-tailored 
chromospheric model calculations to fit  the whole profile of
chromospheric lines in RGB stars of a few globular
clusters. 
Mauas et al. (2006, hereinafter Paper I) studied the Ca II K and
the H$\alpha$ line in RGB stars in NGC2808, and M\'esz\'aros et al. (2009b) 
modelled H$\alpha$ for RGB stars in M13, M15 and M92. 
Although some dependence on luminosity, temperature, and metallicity can be seen, 
the possible transient occurrence of the 
mass loss phenomenon produces considerable uncertainty in the final results. 
  
The aim of the present study is to compare the kinematical structure of the 
chromosphere, i.e. the presence and intensity of a velocity field, in a few 
metal-rich and metal-poor stars with similar characteristics in all other 
respects, to derive an estimate of the mass-loss rate and highlight a possible 
effect of metallicity on the mass-loss phenomenon. A comparison is also  made 
with our previous study on the metal-intermediate globular cluster NGC 2808. 

The stellar system  $\omega$ Cen is the ideal place for this test, as it contains 
multiple stellar populations with different metallicities.

\section{The data }\label{s:data}

\begin{figure}[h]    
\begin{center}     
\includegraphics[width=8cm]{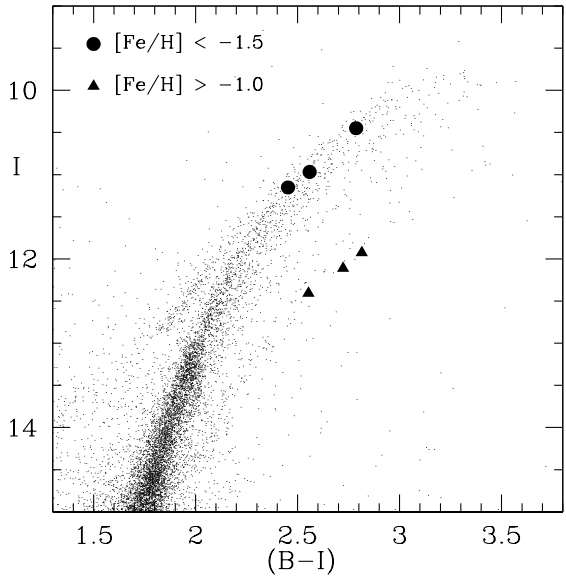}    
\caption{I,(B--I) colour-magnitude diagram of $\omega$ Cen showing the RGB: 
the six targets of the present study are plotted as big filled circles (metal poor) 
and big filled triangles (metal rich). 
} 
\label{f:cmd}      
\end{center}     
\end{figure}

For this study we selected three metal-rich and three metal-poor red giant
stars from the extensive sample studied by Pancino (2003, hereafter P03) 
(see Table \ref{t:stars} and Fig. \ref{f:cmd}). 
The selected stars have bolometric magnitudes between 1 and 2.5 fainter than 
the RGB tip.

All  data were obtained as part of a large observational programme for 
the detailed study of the $\omega$ Cen sub-populations 
(P03; Pancino et al. 2004; Ferraro et al. 2004; Sollima et al. 2005a,b). 

The visual BVI photometry was taken with the WFI at the 2.2m ESO-MPI telescope 
(La Silla, Chile) by P03. 
Infrared photometry was taken with SOFI at the ESO NTT (La Silla, Chile) 
for the initial study of these stars (P03, Sollima et al. 2004),
but in the present analysis we used JK data  from the 2MASS database. 
The photometric data are listed in Table \ref{t:stars}. 

\begin{table*}
\caption{Programme RGB stars in $\omega$ Cen: photometric parameters. Coordinates 
and JK photometry are taken from the 2MASS database, visual photometry is from P03.}
\label{t:stars}
\centering
\begin{tabular}{llccccccc}
\hline\hline
& & & & & & & & \\
Star & LEID & RA & DEC & B & V & I & J & K  \\
\hline
ROA159      & 33011 & 13 25 17.5 & -47 24 26.6 & 13.24 & 11.98  & 10.45 & ~9.537 & ~8.715  \\
ROA256      & 41039 & 13 25 53.8 & -47 28 12.2 & 13.53 & 12.28  & 10.97 & ~9.977 & ~9.190  \\ 
ROA238      & 37247 & 13 27 02.4 & -47 26 06.0 & 13.61 & 12.44  & 11.15 & 10.191 & ~9.363  \\
\\
WFI321293   & 42042 & 13 26 03.8 & -47 28 32.8 & 14.96  & 13.69  & 12.41 & 11.286 & 10.448  \\
ROA523      & 48321 & 13 27 04.8 & -47 31 01.3 & 14.74  & 13.39  & 11.93 & 10.678 & ~9.732  \\
WFI140419   & 32149 & 13 27 14.5 & -47 23 51.2 & 14.83  & 13.49  & 12.11 & 10.960 & 10.072  \\
\hline
\end{tabular}
\end{table*}

\begin{table*}
\begin{center}
\caption{Programme RGB stars in $\omega$ Cen:  physical parameters. The parameters in 
columns 2-5 are taken from P03  (the notation {\em mp} for ROA256 means {\em metal poor}), 
and BC$_K$ are derived from P03 temperatures and the calibration by Montegriffo et al. (1998). 
For the expansion velocities and mass outflow rates in columns 11 and 12 see Sect. 5.1.
}
\label{t:stars_phys}
\centering
\begin{tabular}{lccccccccccc}
\hline\hline
&&&&&&&&&&& \\
Star  & [Fe/H] & [Ca/H] &  $T_{eff}$ & log$g$ & BC$_K$ & M$_{bol}$ & $logL/L_{\odot}$ & $R/R_{\odot}$ & $M/M_{\odot}$ & V$_{exp}$ & $\dot{M}$ \\
      &        &        & $\pm$~100 K  &  $\pm$~0.2 dex & &  &   &        &       &  (km~s$^{-1}$)  & ($M_{\odot}~yr^{-1}$) \\ 
\hline
      &       &        &   &   &        &           &                  &               & &&\\ 
ROA159      & -1.72$^{(1)}$  & -1.37 &  4200 & 0.9 &  2.386 & -2.64 & 2.952 & 56 & 0.88 & ~5 & 1.1 10$^{-9}$ \\
ROA256      & -1.71$^{(1)}$  &  mp   &  4300 & 1.1 &  2.316 & -2.23 & 2.788 & 45 & 0.87 & 15 & 6.0 10$^{-9}$\\ 
ROA238      & -1.80$^{(1)}$  & -1.34 &  4200 & 1.2 &  2.386 & -1.99 & 2.692 & 42 & 0.90 & 15 & 3.2 10$^{-9}$\\
\\
WFI321293   & -0.72          & -0.35 &  4200 & 1.6 &  2.386 & -0.90 & 2.256 & 25 & 0.87 & ... & ... \\
ROA523      & -0.65          & -0.25 &  4200 & 1.4 &  2.386 & -1.62 & 2.544 & 35 & 1.14 & 40 & 5.0 10$^{-10}$\\
WFI140419   & -0.68          & -0.34 &  4200 & 1.5 &  2.386 & -1.28 & 2.408 & 30 & 1.05 & ... & ... \\
\hline
\end{tabular}
\end{center}
Note: $^{(1)}$ - these values are from Cayrel de Strobel et al. (2001).
\end{table*}

The spectroscopic data were taken in April 2001 with UVES in slit mode (R$\sim$45,000) 
at the VLT (ESO-Chile), sampling simultaneously the spectral ranges  350-460 nm 
(Ca II H-K) and  580-700 nm (H$\alpha$) with S/N $\sim$ 100/px. 
The spectra were reduced with the optimal extraction procedure of the UVES pipeline.

\subsubsection{Atmospheric parameters}

The values of metallicity, effective temperature, and gravity  listed in Table \ref{t:stars_phys} 
(columns 2-5) are taken from P03, who derived initial values from the available multiband photometry 
and further refined them through the detailed abundance analysis of the relevant Fe I and Fe II
lines. The three metal-poor stars are also included  in the high-resolution study by 
Cayrel de Strobel et al. (2001), who derived very similar values of metallicity, 
effective temperature, and gravity. 
On the contrary, the low-resolution spectral survey by van Loon et al. (2007)
produced significantly different  results for the four stars that are in common with 
the present study (i.e. ROA159, ROA523, WFI321293 
and WFI140419), but the value of that work lies in its statistical power rather than in the 
accuracy of the individual estimates.   

For a consistency check we estimated the temperatures also from the J-K colours using 
the temperature scale and calibration by Montegriffo et al. (1998), 
and  assuming for 
$\omega$ Cen a reddening E(B-V)=0.11 mag from Lub (2002), and the relations  
E(J-K)=0.52E(B-V) and A$_K$=0.36E(B-V) from Cardelli et al. (1989). 
The results are very close to those of P03 within the errors of these determinations 
(typically $\pm$~100 K). Therefore we used the P03 temperature values to estimate 
the K bolometric corrections BC$_K$ from the calibration by Montegriffo et al. (1998). 
The luminosities were then derived from the K magnitudes assuming for $\omega$ Cen  
a distance modulus (m-M)$_0$=13.70~$\pm$~0.10 mag  (Cacciari et al. 2006), 
and  for the Sun M$_{bol}(\odot)$=4.74~mag.

The stellar radii are obtained from the basic equation of stellar structure 
$$logR/R_{\odot} = 0.5logL/L_{\odot}-2logT_{eff}+7.522$$ 
and the values of mass are estimated from the relation 
$$logM/M_{\odot} = logL/L_{\odot} + logg - 4logT_{eff} + 10.607$$.  
The typical mass for these stars is about 0.8-0.9~M$_{\odot}$ from stellar evolution 
theory, and indeed the values we obtain  agree well  considering the 
errors associated to the temperature and gravity estimates. 
 
The values of the above-mentioned physical parameters are listed in 
Table \ref{t:stars_phys}.

\section{The model chromospheres}\label{s:mods}

To build our models, we adopted mean values of log(g) = 1.5, and metallicities
[Fe/H]=-0.67 for the metal-rich stars. For the metal-poor stars we
used log(g) = 1. and [Fe/H]=-1.6. In both cases we adopted a value of  
[$\alpha$/Fe]=+0.3. These values are in line with those listed in  
Table \ref{t:stars_phys}. 

For each group of stars  we used as a starting point the photospheric
models computed by Kurucz (2005)\footnote{Kurucz, R.L., 2005, 
$http://kurucz.harvard.edu/grids.html$} 
with the closest set of parameters to those of the observed stars. 
For the metal-rich stars we used the model corresponding to T$_{eff}$ = 4250 K,
log(g)=1.5, and [Fe/H] = -0.5, and for the metal-poor stars the model
with T$_{eff}$ = 4250 K, log(g)=1 and [Fe/H] = -1.5. 
We point out that once we adopted a given temperature 
{ \it vs.} metallicity distribution, we recomputed the densities and
all atomic
populations {\it ab initio}. 

On top of the photospheric model, we superimposed a chromospheric rise. We initially 
tried models similar to the ones used in Paper I, with T increasing
linearly with negative log($m$), where $m$ is the mass column density. 
However, we found that we needed to increase T faster just above T$_{min}$ 
and  form a chromospheric plateau to reproduce the Ca II K profiles. 
Although these models are therefore much more complex than the ones in Paper I, 
they resemble more closely those usually used for dwarf stars and, in particular, 
for the Sun.

Upon the chromosphere, we added a ``transition region'' where the  
temperature rises abruptly up to $1.5 \times 10^5$ K, to assure
convergence of the calculations. The structure of this ``transition
region'' has no influence on the computed profiles. 

Since the asymmetries observed are not very large, they can be
considered as a second order change in the line profiles. Furthermore,
calculations including velocity fields take much longer to compute. 
Therefore, we first found a reasonable agreement between the computed
profiles and the observed ones, and then we included a velocity field
to reproduce the asymmetries.  We modified this 
velocity field until a satisfactory match between observed and computed 
profiles was found. The advantage of this approach is that the velocities are 
not ``measured'' from the profiles, but modelled self-consistently along with 
the rest of the atmospheric parameters, and the region corresponding to each 
value of the velocity field is well-determined, unlike for other methods such 
as the bisector.

For every $T$ vs. log($m$) distribution we used the programme Pandora 
(Avrett \& Loeser 1984) to solve the non-LTE radiative
transfer and the statistical and hydrostatic equilibrium equations,
assuming turbulent velocities varying from 2 to 20 km~s$^{-1}$.
We self-consistently computed non-LTE populations for 10 levels
of H 
(for details on the atomic models, see Falchi \& Mauas 2002), 13 of He {\sc i} 
(Mauas et al. 2005), 
9 of C {\sc i} 
(Mauas, Avrett \& Loeser 1989), 
15 of Fe {\sc i}, 8 of Si {\sc i} 
(Cincunegui \& Mauas 2001),  
8 of Ca {\sc i} and Na {\sc i},  6 of Al {\sc i} 
(Mauas, Fern\'andez Borda \& Luoni 2002), 
and 7 of Mg {\sc i} 
(Mauas, Avrett \& Loeser 1988). 
In addition, we computed 6 levels of He {\sc ii} and Mg {\sc ii}, 
and 5 of Ca {\sc ii}. We also self-consistently computed the
contribution to the opacity of the most abundant molecule, CO 
(Mauas et al. 1990).

\begin{figure}[h]    
\begin{center}     
\includegraphics[width=8cm]{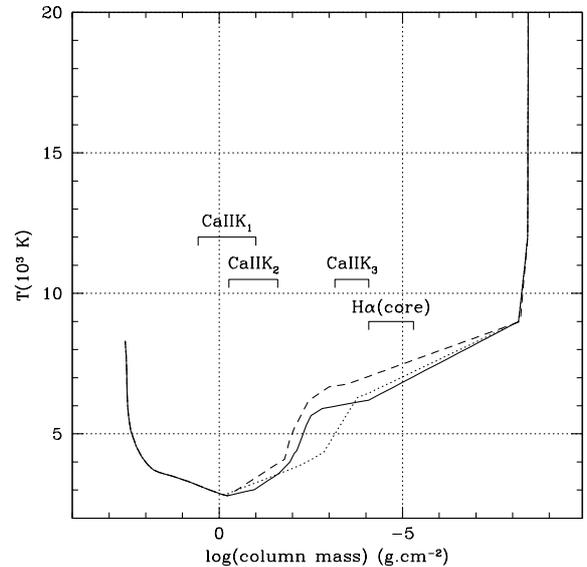}      
\caption{Our chromospheric models for the metal-rich stars: ROA 523
  (solid line), WFI 140419 (dotted line) and WFI 321923 (dashed
  line). The approximate depth of formation of the H$\alpha$ core 
  and of the Ca {\sc ii} K$_1$, K$_2$ and K$_3$ line components
  are also indicated.  } 
\label{f:mod_rich}      
\end{center}     
\end{figure}      

For every species considered we included all bound-free 
transitions and the most important bound-bound transitions. Ly-$\alpha$, the
Ca {\sc ii} H and K, and Mg {\sc ii} $h$ and $k$ lines were all 
computed with a full partial-redistribution treatment 
(Falchi \& Mauas 1998, for a discussion). 

In our calculations we assumed a plane-parallel atmosphere, for simplicity. 
However, once the final model was obtained, we computed the emitted profiles 
assuming a spherically symmetric atmosphere, and we found no significant 
changes in the emitted profiles with respect to the plane-parallel 
approximation.

\section{Analysis and results}\label{s:anres}

\begin{figure}[h]      
\begin{center}     
\includegraphics[width=8cm]{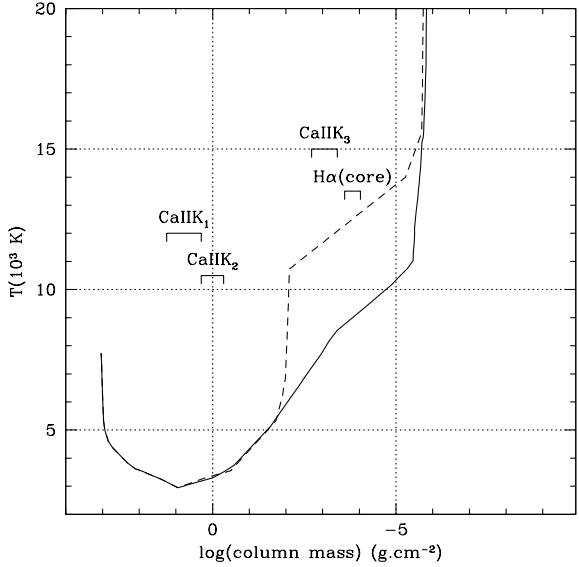}       
\caption{Same as Fig. \ref{f:mod_rich} for the metal-poor stars: ROA159
  (dashed line), ROA238 and ROA256 (solid line).}
\label{f:mod_poor}      
\end{center}     
\end{figure}

\begin{figure}[h]      
\begin{center}     
\includegraphics[width=8cm]{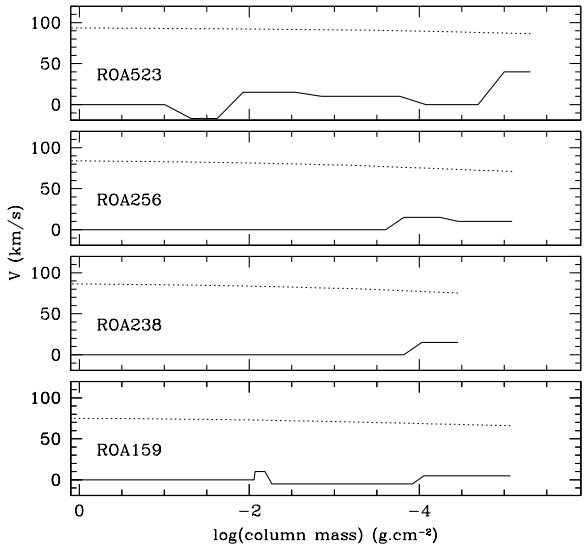}       
\caption{Velocity as a function of depth (solid line) for 
the metal-rich star ROA523 (top panel) and the three metal-poor stars 
(lower panels). A positive velocity implies outflow. 
No velocity fields were required for the two 
metal-rich stars WFI321293 and WFI140419, which are therefore not 
shown in this plot. The velocities were derived from the best match
with the Ca K and H$\alpha$ lines and are compared to 
the corresponding escape velocity (dotted line).   }
\label{f:velos}      
\end{center}     
\end{figure}

In Fig. \ref{f:mod_rich} we show the atmospheric models for the metal-rich 
stars with the depth of formation of the lines we used as diagnostics. 
The depth of formation is defined as the depth where
most of the observed photons are formed. 
The models for the metal-poor stars are shown in 
Fig. \ref{f:mod_poor}. The velocities that give the best fit with the
observed line asymmetries are plotted in Fig. \ref{f:velos}. For two of
the metal-rich stars, WFI 140419 and WFI 321293, the observed profiles
are symmetric, and we therefore did not need to include any velocities
to reproduce them. 

\begin{figure*}[ht]      
\begin{center}     
\includegraphics[width=15.0cm]{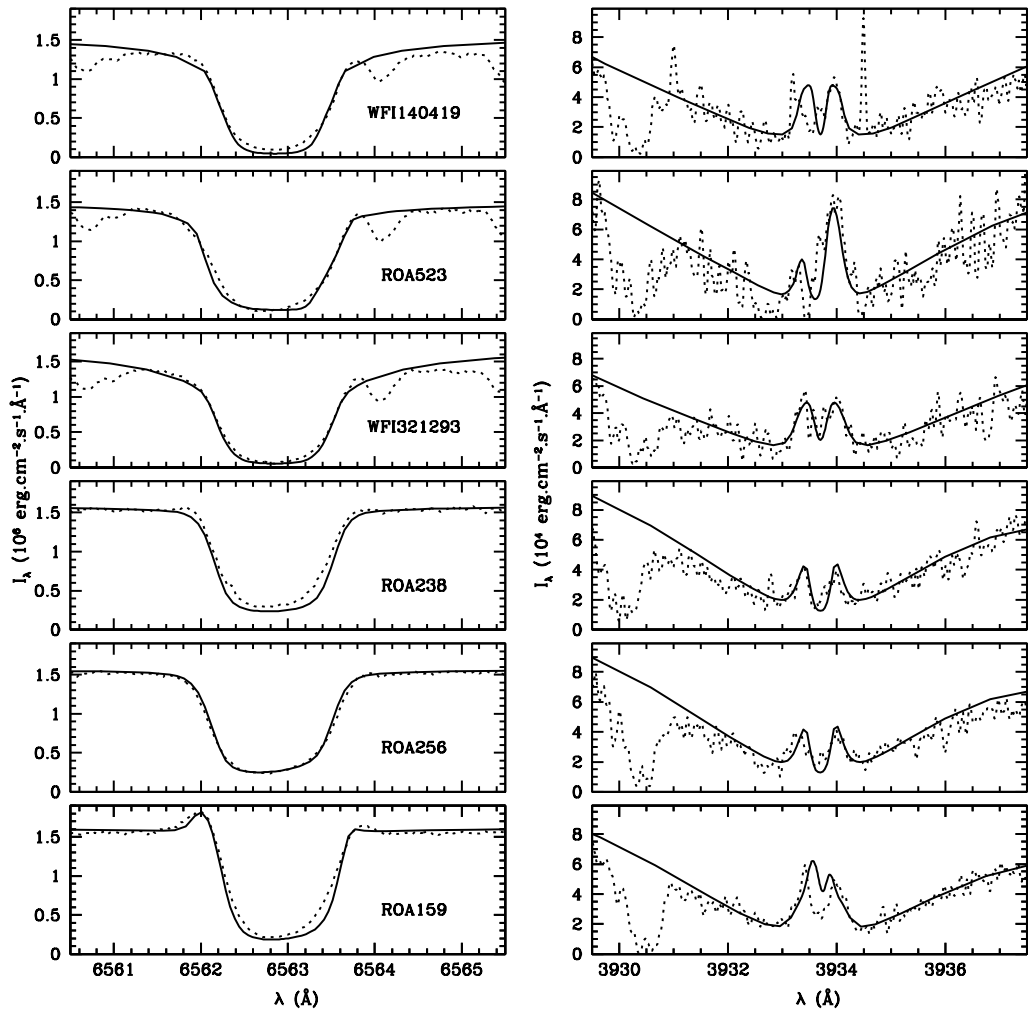}      
\caption{For every star, the observed profiles (dotted lines) are
  compared with the synthetic profiles computed with the velocity
  fields shown in   Figs. \ref{f:velos} (solid lines). The
  profiles for H$\alpha$ (left panel) and Ca {\sc ii} K (right panel)
  for the metal-rich stars are shown in the three uppermost panels,
  and those for the metal-poor stars in the three lowermost ones.}
\label{f:all_prof}      
\end{center}        
\end{figure*}

The computed H$\alpha$ and Ca {\sc ii} K line
profiles are compared with the observations in Fig. \ref{f:all_prof}.
The observed profiles have been 
normalized to the continuum emission implied by Kurucz's model, and shifted by 
a quantity corresponding to the individual stellar radial velocity to 
move the profiles to their rest wavelengths. 

One can see in the figure that the agreement between computed 
and observed profiles is quite good. In Paper I, the computed
H$\alpha$ profiles were broader and slightly deeper in the line core
than the observed ones. However, this discrepancy is much smaller in
the present models. 

Regarding the Ca {\sc ii} profiles, they are much noisier and are
therefore more difficult to interpret. In a couple of cases, the models
produced higher K$_3$ central intensities than the observed profiles,
as was generally the case in Paper I. As we discussed in that paper,
this can be due to the assumption
of a homogeneous chromosphere, since any contribution of a pure photospheric
line should be noticed mainly in the line centre.   
We do not expect a significant contribution of interstellar Ca, since 
we do not see evidence of it in our spectra, and it could contribute 
to the Ca {\sc ii} core profile only if the velocity of the interstellar
medium were similar to that of the cluster ($\sim$ 230 km~s$^{-1}$).

However, on average the match to the observations given by these
models is better than the one we found in Paper I. This can be 
because these stars show lower velocities, and in only one
case there is a noticeable emission in the H$\alpha$ wings. On the
other hand, this may also be because these models
are much more complex, since here we did not require a linear T vs.
log(m). 

Regarding the models, there is a large difference between those for
the metal-rich and metal-poor stars, with the latter ones starting at
smaller column masses. On the other hand, the three models for the
metal-rich stars agree with each other. Indeed, 
the computed profiles for these stars are unaffected by the
structure of the regions above log(m)$=10^{-7}$ (see Fig. \ref{f:mod_rich}),
and we therefore used similar structures for each model in that region.

For the metal-poor stars, the model for ROA238 and ROA256 is the same,
and the profiles differ only because of different velocity fields. The model
for ROA159 is much hotter. This  is reflected
in the emission wings in the H$\alpha$ profile, which, as in the
models of Paper I, is a direct sign of a steep chromospheric
temperature rise. 

\subsection{Velocity fields and mass loss}

In general, the stars studied here have smaller velocity fields than those 
modelled in Paper I. Furthermore, in two
cases (WFI140419 and WFI321293), we did not need to include velocities
to fit the observations. For the other stars, we show the full
velocity field in Fig. \ref{f:velos}. For comparison, we 
also show the corresponding escape velocity at each height 
Z from the centre of the star, computed as V$_{esc}$=620\ (M/Z)$^{1/2}$,
where Z and the stellar mass M are in solar units,  and the 
velocity is in km s$^{-1}$. 
We assumed a typical mass M=0.85 M$_{\odot}$ for all stars. 

The H$\alpha$ line core is the feature that forms higher up
in the atmosphere, so we can only infer the velocity up to the 
height where it is formed, about 20-30\% 
of a stellar radius above the photosphere. 
All metal-poor stars show low expansion velocities 
($\le$ 15 km~s$^{-1}$) at the most external point, and only 
the metal-rich ROA523 has a high expansion velocity (40 km~s$^{-1}$). 
These values are similar to those obtained in Paper I. 
However, since the gravities of the present targets are 5-10 times 
larger than those in Paper I, the escape velocities are
higher too and we cannot be certain that  the material
above this level is indeed escaping from the atmosphere.

However, there is a mass outflow at this height, which can be
estimated using the simple formulation
$$\dot{M} = 1.3 \times m_H \times N_H \times 4\pi
  \times R^2 \times V\ $$,
  
\noindent where $m_H$ and $N_H$ are the mass and the density of
hydrogen atoms, R  the distance from the stellar centre, 
and V  the velocity of the outermost layer for which we
have a determination. The factor 1.3 takes care of the helium
abundance Y=0.23.  
The expansion velocities detected in the outermost part of the atmosphere 
where the H$\alpha$ core forms are shown in Table \ref{t:stars_phys}, 
as well as the mass outflow rates calculated from the above equation. 
 
As there are no signs of a reversal in the velocity field, 
{\it i.e.} there is no material falling back in, 
the most probable situation is that the velocity continues to
increase with distance, until eventually it reaches the escape 
velocity which, of course, is decreasing with distance.

\subsection{Energy requirements} 

To estimate the energy requirements to sustain the
chromosphere, {\it i.e.} to constrain the possible heating
mechanisms at work, we
compute the radiative cooling rate $\Phi$ (erg cm$^{-3}$ s$^{-1}$),
namely the net amount of energy radiated at a given depth by the
atmosphere, which is given by

$$ \Phi = 4 \pi \int \kappa_{\nu}\ (S_\nu-J_\nu)\ d\nu$$. 

A positive value implies a net loss of energy 
(cooling), and a negative value represents a net energy absorption 
(heating).

\begin{figure}[h]  
\begin{center}     
\includegraphics[width=8cm]{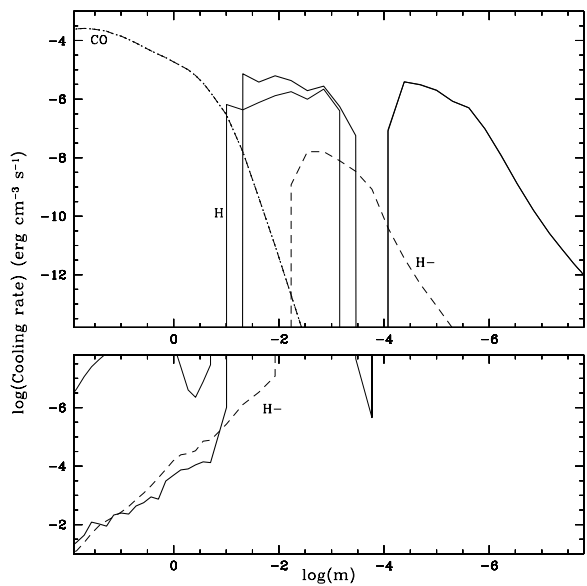}       
\caption{Log$ \Phi$, {\it i.e.} the total net radiative cooling rate, as a 
function of depth for our model for the star ROA523 (thick line), and
the most important  contributions to it.  In the upper panel we show
positive values ({\it i.e.} net cooling), and in the lower panel we
show negative values ({\it i.e.} net heating).
}
\label{f:cool}      
\end{center}     
\end{figure}

In this study we compute the contributions due to H$^{-}$,
H, He {\sc i}, Mg {\sc i} and {\sc ii}, 
Ca {\sc i} and {\sc ii}, Fe {\sc i}, Si {\sc i}, Na {\sc i},
Al {\sc i}, and CO. The overall results and the
most important individual contributions for the model that represents
star ROA523 are presented in Fig. \ref{f:cool}. We do not include the
rates for the other models, which are very similar.
                      
The main difference with the results obtained for other stars (e.g. in
Paper I) is the presence of a negative region in the
mid-chromosphere, caused by backwarming by Ly-$\alpha$ emission, which has 
its origin in a more external layer. 
In the rest of the atmosphere, these rates are
similar to those obtained in Paper I. As we pointed out there, the
energy required is much lower for giant stars than for dwarf stars of
similar T$_{eff}$ (Mauas et al. 1997; Vieytes et al. 2005, 2009).
Indeed, the cooling rate
reaches up to 10$^{-4}$ erg cm$^{-3}$ s$^{-1}$ for these giant stars, 5
orders of magnitude smaller than for the dwarfs. On the other hand, 
the energy {\it per particle} is of the same order of magnitude, since
the particle density is also 5 orders of magnitude smaller for the
present stars.

As in the models in Paper I, the cooling rate in the coolest parts of
the atmosphere is negative and due almost entirely to H$^-$,  as was
already noted for dwarf stars. However, almost all cooling in
that region is due to CO, which is the only molecule included in the
calculations. Because at these low temperatures the presence of other
molecules should be expected, it is reasonable to expect that the
inclusion of other molecular species in the calculations should bring
the atmosphere closer to radiative balance. 

In the mid- and high-chromosphere the energy balance is 
determined essentially by the hydrogen cooling rate. This  is of
course more important for the metal-poor stars, where the influences
of cooling by metallic lines such as the Mg {\sc ii} and Ca {\sc ii} can
be neglected.
 
\section{Summary and conclusions}\label{s:conc} 

We studied three metal-rich and three metal-poor red giant 
stars in the stellar cluster $\omega$ Cen. To analyse in detail 
and characterize the possible outward velocity fields that are indicative
of mass outflow, we built model chromospheres and computed synthetic
line profiles that we compared with high-resolution profiles of the
H$\alpha$ and Ca {\sc ii} K lines, unlike M\'esz\'aros et al. (2009b),  
who only used H$\alpha$.

We can summarize our results as follows: 

\begin{itemize}

\item
As we noticed in Paper I, 
the existence of emission in the Ca {\sc ii} K lines
is of chromospheric origin and does not require the presence of  
a velocity field. This is also true for the H$\alpha$ line.
However, a fairly steep rise in temperature with log(m) is needed 
to explain this emission, as can be seen in the model for ROA159.

\item
On the other hand, the asymmetries in the Ca {\sc ii} K and H$\alpha$
line profiles indicate the presence of velocity fields 
in four out of six stars.  
These ``expansion'' velocities are less than about 15 km~s$^{-1}$ 
except for ROA523,  where it reaches about  40 km~s$^{-1}$, they are all 
outward and provide clear evidence that some mass outflow occurs in 
these stars.  

\item 
The only metal-rich star showing outward velocity and mass outflow is 
ROA523, the brightest of its group. On the other hand, all the metal-poor 
stars are brighter than the metal-rich ones, and show evidence of outward 
velocities and slightly higher values of mass outflow rates. 
This suggests that there is a stronger correlation of the mass loss 
phenomenon with luminosity rather than with metallicity, as found in 
several previous studies (see Sect. 1).       

\item
On the assumption that the mass outflow eventually escapes the star, 
the rates of mass loss are estimated as a few 
10$^{-9}$-10$^{-10}$ M$_{\odot}$ yr$^{-1}$, in general agreement with
previous estimates from the Mg {\sc ii} $k$ line 
(Dupree et al. 1990a,b, 1994; Smith \& Dupree 1998)
and with the requirements of the stellar evolution theory.  

\item In Paper I we obtained a similar result, namely four stars with 
low outward velocity  (less than 20 km~s$^{-1}$) and one with a velocity of 
about 50 km~s$^{-1}$. The globular cluster NGC 2808 has [Fe/H] $\sim$ -1.15,  
which is intermediate between the two populations studied here, and all  
stars studied in Paper I have $logL/L_{\odot} >$~3.0 and somewhat 
larger radii than the present targets.
The mass-loss rates estimated in Paper I are quite similar to the 
present ones, supporting the suggestion that a higher luminosity   
could favour mass loss. However, the likely transient nature of the 
mass-loss  phenomenon may contribute to mask any correlation with 
metallicity and luminosity.

\item
As in Paper I, the energy per unit volume required to sustain the chromosphere 
is much smaller than the energy needed for a dwarf star with similar 
T$_{eff}$. However, the energy {\it per particle} is of similar amount.

\end{itemize}

As a final conclusion, this study provides additional evidence that outward 
velocity fields and mass motions exist along the RGB at fainter levels 
(i.e. as much as 2.5 mag) than the RGB tip, as already suggested by 
Origlia et al. (2010) based on evidence for ``historic'' mass loss 
from dust emission in 47 Tuc.

\begin{acknowledgements} 

PM and CC acknowledge the support of a Visitor Exchange Program within the bilateral 
agreement between the Ministerio de Asuntos Exteriores (Argentina) and the Ministero 
degli Affari Esteri (Italy). PM acknowledges the EADIC consortium of the European Comission 
for a travel grant to the University of Bologna. 

\end{acknowledgements}


\begin{thebibliography}{}
\bibitem[2008]{and08} Andretta, V., {Mauas},
  P.~J.~D., {Falchi}, A. \& {Teriaca}, L. 2008, \apj, 681, 650
\bibitem[1984]{av84} Avrett E.H., \& Loeser R., 1984, in {\it Methods in 
   Radiative Transfer}, ed. W. Kalkofen, Cambridge, Univ. Press, p. 341.
\bibitem[2009]{bar09} Barmby, P., Boyer, M.L., Woodward, C.E., Gehrz, R.D., 
van Loon, J.Th., Fazio, G.G., Marengo, M. \& Polomski, E. 2009, \aj, 137, 207
\bibitem[1990]{bat90} Bates, B., Catney, M.G. \& Keenan, F.P. 1990, \mnras,     
   245, 238     
\bibitem[1993]{bat93} Bates, B., Kemp, S.N. \& Montgomery, A.S. 1993,   
 \aaps, 97, 937      
\bibitem[2006]{boy06} Boyer, M. L., Woodward, C. E., van Loon, J. Th., Gordon, 
K. D., Evans, A., Gehrz, R. D., Helton, L. A., \& Polomski, E. F. 2006, \aj, 132, 1415
\bibitem[2008]{boy08} Boyer, M. L., McDonald, I., van Loon, J.Th., Woodward, C. E., 
Gehrz, R. D., Evans, A. \& Dupree, A.K. 2008, \aj, 135, 1395
\bibitem[2009]{boy09} Boyer, M. L., McDonald, I., van Loon, J.Th., 
   Gordon, K.D. et al. 2009, \apj, 705, 746
\bibitem[2010]{boy10} Boyer, M. L., van Loon, J. Th., McDonald, I. et al. 
     2010, \apj, 711, L99 
\bibitem[1983]{cf83} Cacciari, C. \&  Freeman, K.C. 1983, \apj, 268, 185
\bibitem[2004]{cac04} Cacciari, C., Bragaglia, A., Rossetti, E. et al. 2004, 
   \aap, 413, 343
\bibitem[2006]{cac06} Cacciari, C., Sollima, A. \& Ferraro, F.R. 2006, 
     \memsai, 77, 245
\bibitem[1989]{ccm89} Cardelli, J.A., Clayton, G.C. \& Mathis, J.S. 1989, 
   \apj, 345, 245
\bibitem[1968]{cr68} Castellani, V. \& Renzini, A. 1968, \apss, 2, 310
\bibitem[Catelan2000]{cat00} Catelan, M. 2000, \apj, 531, 826
\bibitem[2001]{cay01} Cayrel de Strobel, G., Soubiran, C. \& Ralite, N. 
   2001, \aap, 373, 159 
\bibitem[2001]{cm01} Cincunegui, C. \& Mauas, P.J.D. 2001, ApJ 552, 877
\bibitem[1976]{coh76} Cohen, J.G. 1976, \apj, 203, L127     
\bibitem[1978]{coh78} Cohen, J.G. 1978, \apj, 223, 487     
\bibitem[1979]{coh79} Cohen, J.G. 1979, \apj, 231, 751     
\bibitem[1980]{coh80} Cohen, J.G. 1980, \apj, 241, 981     
\bibitem[1981]{coh81} Cohen, J.G. 1981, \apj, 247, 869     
\bibitem[1996]{dc96} D'Cruz, N.L., Dorman, B., Rood, R.T. \& O'Connell, R.W.      
   1996, \apj, 466, 359 
\bibitem[1984]{dup84} Dupree, A.K., Hartmann, L. \& Avrett, E.H. 1984,      
  \apj, 281, L37 
\bibitem[1990a]{dup90a} Dupree, A.K., Harper, G.M., Hartmann, L., Jordan, C., 
   Rodgers, A.K. \& Smith, G.H. 1990a, \apj, 361, L9 
\bibitem[1990b]{dup90b} Dupree, A.K., Hartmann, L. \& Smith, G.H. 1990b, 
   \apj, 353, 623
\bibitem[1992]{dup92} Dupree, A.K., Sasselov, D.D \& Lester, J.B. 1992, 
   \apj, 387, L85 
\bibitem[1994]{dup94} Dupree, A.K., Hartmann, L., Smith, G.H., Rodgers, A.W.,      
   Roberts, W.H. \& Zucker, D.B. 1994,  \apj, 421, 542  
\bibitem[1995]{dup95} Dupree, A.K. \& Smith, G.H. 1995, \aj, 110,  405 
\bibitem[2009]{dup09} Dupree, A.K., Smith, G.H. \& Strader, J. 2009, \aj, 138, 1485
\bibitem[2003]{eva03} Evans, A., Stickel, M., van Loon, J.Th., Eyres, S.P.S.,
   Hopwood, M.E.L. \& Penny, A.J. 2003, \aap, 408, L9
\bibitem[2008]{fab08} Fabbri, S., Origlia, L., Rood, R.T., Ferraro, F.R., 
   Fusi Pecci, F. \& Rich, M. 2008, Mem. S.A.It. Vol. 79, 720
\bibitem[1998]{fm98} Falchi, A. \& Mauas, P.J. 1998, \aap, 336, 281
\bibitem[2002]{fm02} Falchi, A. \& Mauas, P.J. 2002, \aap, 387, 678 
\bibitem[2004]{fer04} Ferraro, F.R., Sollima, A., Pancino, E., Bellazzini, M., 
   Straniero, O., Origlia, L. \& Cool, A.M. 2004, \apj, 603, L81
\bibitem[2001]{fre01} Freire, P.C., Kramer, M., Lyne, A.G., Camilo, F., 
   Manchester, R.N. \& D'Amico, N. 2001, \apjl, 557, 105 
\bibitem[1988]{fro88} Frogel, J.A. \& Elias, J.H. 1988, \apj, 324, 823
\bibitem[1993]{ffp93} Fusi Pecci, F., Ferraro, F.R., Bellazzini, M.,      
   Djorgovski, G.S., Piotto, G. \& Buonanno, R. 1993, \aj, 105, 1145
\bibitem[1988]{gil88} Gillet, F.C., deJong, T., Neugebauer, G., Rice, W.L. 
\& Emerson, J.P. 1988, \aj, 96, 116
\bibitem[1984]{gra84} Gratton, R.G., Pilachowski, C.A. \& Sneden, C. 1984,      
   \aap, 132, 11  
\bibitem[2007]{kal07} Kalirai, J.S., Bergeron, P., Hansen, B.M.S., Kelson,
D.D., Reitzel, D.B., Rich, R.M. \& Richer, H.B. 2007, \apj, 671, 748 
\bibitem[2007]{vl07} van Loon, J.Th., van Leeuwen, F., Smalley, B., Smith, A.W., 
   Lyons, N.A., McDonald, I. \& Boyer, M.L. 2007, \mnras, 382, 1353
\bibitem[2002]{lub02} Lub, J. 2002, in ``Omega Centauri, a unique window into 
Astrophysics'', ASP Conf. Ser. 265, 95
\bibitem[1996]{lyo96} Lyons, M.A., Kemp, S.N., Bates, B. \& Shaw, C.R. 1996,      
   \mnras, 280, 835      
\bibitem[1978]{mp78} Mallia, E.A. \&; Pagel, B.E. 1978, \mnras, 184, 55P
\bibitem[1988]{mau88} Mauas, P.J.D., Avrett, E.H. \& Loeser, R. 1988, 
   \apj, 330, 1008.
\bibitem[1989]{mau89} Mauas, P.J.D., Avrett, E.H. \& Loeser, R. 1989, \apj, 345, 1104
\bibitem[1990]{mau90} Mauas, P.J.D., Avrett, E.H. \& Loeser, R. 1990, \apj, 357, 279
\bibitem[1997]{mau97} Mauas, P. J. D., Falchi, A., Pasquini, L., \& Pallavicini, R. 1997, 
   \aap, 326, 249
\bibitem[2002]{mau02} Mauas, P.J.D., Fern\'andez Borda, R. \& Luoni, M.L. 2002, 
   \apjs, 142, 285
\bibitem[2005]{mau05} Mauas, P.J.D., {Andretta}, V.,
  {Falchi}, A.,{Falciani}, R., {Teriaca}, L. \& {Cauzzi}, G. 2005, \apj, 619, 604
\bibitem[2006]{mau06} Mauas, P.J.D., Cacciari, C. \& Pasquini, L. 2006, 
   \aap, 454, 609 (Paper I) 
\bibitem[2009]{mcd09} McDonald, I., van Loon, J.Th., Decin, L., 
   Boyer, M.L., Dupree, A,.K., Evans, A., Gehrz, R.D. \& Woodward, C.E. 2009, \mnras, 394, 831
\bibitem[2008]{mes08} M\'esz\'aros, Sz., Dupree, A.K., \& 
   Szentgyorgyi, A.H. 2008, \aj, 135, 1117
\bibitem[2009a]{mes09a} M\'esz\'aros, Sz., Dupree, A.K., \& 
   Szalai, T. 2009a, \aj, 137, 4282
\bibitem[2009b]{mes09b} M\'esz\'aros, Sz., Avrett, E.H. \& Dupree, A.K.
   2009b, \aj, 138, 615   
\bibitem[1998]{mon98} Montegriffo, P., Ferraro, F.R., Origlia, L. 
   \& Fusi Pecci, F. 1998, \mnras, 297, 872   
\bibitem[1996]{ori96} Origlia, L., Ferraro, F.R. \& Fusi Pecci, F. 1996, 
   \mnras, 280, 572
\bibitem[1997]{ori97} Origlia, L., Scaltriti, F., Anderlucci, E., Ferraro,
   F.R. \& Fusi Pecci, F. 1997, \mnras, 292, 753
\bibitem[2002]{ori02} Origlia, L., Ferraro, F.R., Fusi Pecci, F. \&  
   Rood, R.T. 2002, \apj, 571, 458
\bibitem[2007]{ori07} Origlia, L., Rood, R.T., Fabbri, S., Ferraro, F.R., 
   Fusi Pecci, F. \&  Rich, R.M. 2007, \apj, 667, 85
\bibitem[2008]{ori08} Origlia, L. 2008, \memsai, 79, 432
\bibitem[2010]{ori10} Origlia, L., Rood, R.T., Fabbri, S., Ferraro, F.R., 
   Fusi Pecci, F., Rich, R.M. \& Dalessandro, E. 2010, \apj, 718, 522
\bibitem[2003]{pan03} Pancino, E. 2003, PhD Thesis, University of Bologna  (P03)
\bibitem[2004]{pan04} Pancino, E., Ferraro, F.R., Bellazzini, M., Piotto, G. 
   \& Zoccali, M. 2004, \apj, 534, L83 
\bibitem[1981]{pet81} Peterson, R.C. 1981, \apj, 248, L31      
\bibitem[1982]{pet82} Peterson, R.C. 1982, \apj, 258, 499 
\bibitem[1975a]{rei75a} Reimers, D. 1975a, in {\it Problems in Stellar Atmospheres and 
Envelopes}, ed. B. Baschek, W.H. Kegel, \& G. Traving (Berlin: Springer), 229
\bibitem[1975b]{rei75b} Reimers, D. 1975b, Mem. Soc. R. Sci. Liege, 8, 369
\bibitem[1973]{rtr73} Rood, R.T. 1973, \apj, 184, 815
\bibitem[1990]{smi90} Smith, G.H., Wood, P.R., Faulkner, D.J. \& Wright, A.E.      
   1990, \apj, 353, 168 
\bibitem[1992]{smi92} Smith, G.H., Dupree, A.K. \& Churchill, C.W. 1992, \aj,      
   104, 2005 
\bibitem[1998]{smi98} Smith, G.H. \& Dupree, A.K. 1998, \aj, 116, 931
\bibitem[2004]{smi04} Smith, G.H., Dupree, A.K. \&+ Strader, J. 2004, \pasp, 
   116, 819 
\bibitem[2004]{sol04} Sollima, A., Ferraro, F.R., Origlia, L., Pancino, E. 
   \& Bellazzini, M. 2004, \aap, 420, 173
\bibitem[2005a]{sol05a} Sollima, A., Ferraro, F.R., Pancino, E. \& Bellazzini, M. 
   2005a, \mnras, 357, 265
\bibitem[2005b]{sol05b} Sollima, A., Pancino, E., Ferraro, F.R., Bellazzini, M., 
   Straniero, O. \& Pasquini, L. 2005b, \apj, 634, 332  
\bibitem[2005]{vie05} 	Vieytes, M., Mauas, P.J.D. \&
  Cincunegui, C. 2005, \aap, 441, 701
\bibitem[2009]{vie09} 	Vieytes, M., Mauas, P.J.D. \&
  Díaz, R.F. 2009, \mnras, 398, 1495
\end{thebibliography}
\end{document}